\documentstyle [12pt] {article}

\newcommand{\be}{\begin{equation}}
\newcommand{\ee}{\end{equation}}
\newcommand{\bea}{\begin{eqnarray}}
\newcommand{\eea}{\end{eqnarray}}
\newcommand{\ba}{\begin{array}}
\newcommand{\ea}{\end{array}}

\newcommand{\tb}{\bar{\tau}}
\newcommand{\ds}{\displaystyle}

\newcommand{\nc}{\newcommand}

\nc{\btu}{\bigtriangleup}
\nc{\im}{{\rm Im}\ }
\nc{\nn}{\nonumber}
\nc{\cd}{\cdot}
\nc{\kae}{K\"ahler\ }
\nc{\zd}{{\bf Z}$_3$\ }
\nc{\te}{$\Theta$\ }
\nc{\vh}{\hat{v}}
\nc{\vt}{\tilde{v}}
\begin{document}

\begin{titlepage}
\begin{flushright}MPI--Ph/93--48 \\ TUM--TH--158/93 \\ hep-th/9307171 \\ \ \\
July 1993
\end{flushright}
\vfill

\begin{center}
{\Large \bf String Unification and Threshold Corrections$^\ast$}   \\
\vskip 1.2cm
{\large \bf P. Mayr}\ , {\large \bf H.P. Nilles}
 \ and\ \ {\large \bf S. Stieberger} \\
\vskip .75cm

{\em Physik Department} \\
{\em Institut f\"{u}r Theoretische Physik} \\
{\em Technische Universit\"at M\"unchen} \\
{\em D--85748 Garching, Germany}
\vskip .25cm
{\rm and}\\
\vskip .25cm
{\em Max--Planck--Institut f\"ur Physik} \\
{\em ---Werner--Heisenberg--Institut---}\\
{\em P.O. Box 401212}\\
{\em D--80805 M\"{u}nchen, Germany}
\end{center}
\vfill
\vspace{1cm}

\begin{center}
{\bf ABSTRACT}
\end{center}

The interpretation of the apparent unification of gauge couplings
within supersymmetric theories depends on uncertainties induced
through heavy particle thresholds. While in standard grand unified
theories these effects can be estimated easily, the corresponding
calculations are quite complicated in string unified theories and do
exist only in models with unbroken $E_6$. We present results for
heavy particle thresholds in more realistic models with gauge group
$SU(3)\times SU(2)\times U(1)$. Effects of Wilson line background fields
as well as the universal part of the (rather mild) threshold corrections
indicate a strong model dependence. We discuss the consequences of our
results for the idea of string unification without a grand unified
gauge group.

\vskip 5mm \vskip0.5cm
\hrule width 5.cm \vskip 1.mm
{\small\small
\noindent $^\ast$ Partially supported by the Deutsche Forschungsgemeinschaft
and the EC under contract no. SC1--CT92--0789.}
\normalsize


\end{titlepage}

An appealing concept for the extension of
the successful standard model (SM) to higher
energy scales is the idea of grand unified theories (GUTS). They provide
a useful tool to gain knowledge about new physics up to the unification
scale $M_X$. The evolution of the gauge couplings according to
the renormalization group equations (RGE) is
determined by the particle spectrum
below $M_X$; therefore the gauge couplings are
sensitive detectors for the presence of each  particle with
nontrivial gauge quantum numbers.
In the context of the minimal supersymmetric extension of the
standard model (MSSM) the extrapolation of the low-energy data for
the $SU(3)\times SU(2)\times U(1)$ gauge couplings is consistent
with grand unification at a scale of  $M_X \approx 2 \cdot 10^{16}\
GeV$ \cite{uni00}. This agreement could serve as an argument for the
existence of a unified gauge group as e.g. $SU(5)$, $SO(10)$ or
$E_6$ above the unification scale $M_X$.

Unfortunately the reliability of this result is weakened by the presence
of uncertainties due to threshold effects at large scales \cite{lan93}.
In the
presence of large group representations, e.g. a $\underline{75}$ in
$SU(5)$ models \cite{jap93},
these corrections can acquire a considerable size and
disturb seriously this successful picture. Moreover there are deep
conceptual problems peculiar to supersymmetric GUTS: the avoidance of
too fast proton decay raises the well--known
problem of doublet--triplet splitting
leading to a revival of the mass hierarchy problem whose removal was
one of the primary
motivations for the introduction of
supersymmetry. In addition the boundary conditions
for the Yukawa couplings at $M_X$ forced by the group structure
put serious constraints on the fermion mass spectrum.
 Finally these theories are to be embedded in a more complete theory
at the natural high energy scale $M_{Planck}$ where gravitational
quantum effects can no longer be neglected.

Amazingly superstring theories, the only known theories which
consistently unify all known interactions including gravity, show
likewise the feature of gauge coupling unification. At the
string scale $M_{string} \approx 0.7 \cdot g_{string} \cdot 10^{18}\
GeV$ \cite{kap88,der91}
the gauge coupling constants of the various group factors are related to
the gravitational coupling constant $G_N$ at tree level as follows
\cite{gin87}:
\be
g_a^2k_a = 4\pi \alpha^{\prime -1}G_N = g_{string}\ , \; \; \forall\;  a \ ,
\ee
where $\alpha^{\prime -1/2}$ is the string tension and $k_a$ is the
Kac--Moody level of the group factor labeled by $a$.
Below $M_{string}$ effective theories are an appropriate description of the
low--energy physics and the couplings begin to evolve according to the RGE as
in ordinary field theories. At the one--loop level the above equation
is modified to \cite{wei80,kap88}
\be
\frac{16\pi^2}{g^2_a(\mu)} = k_a \frac{16\pi^2}{g_{string}^2} +
b_a \ln \left( \frac{M^2_{string}}{\mu^2}\right) + \btu_a \ ,
\ee
where $b_a$ are the coefficients of the $\beta$--function via
$\beta_a = b_ag_a^3/16\pi^2$.

String theories in four dimensions do not necessarily require the presence
of a grand unified gauge group at $M_X$. This fact could be used for the
most elegant solution of the so-called doublet-triplet splitting
(or missing partner mechanism): there is no need for the partner
triplet to exist.
In addition the constraints on
the Yukawa couplings need not be imposed.
On the other hand the
unification scale $M_X$ is now shifted to $M_{string}$ and therefore
differs by a factor of approximately
25 from the extrapolation of the gauge couplings
according to the MSSM particle content. Again there are uncertainties due
to thresholds corrections.
Possibly they might resolve
the discrepancy between $M_X$ and $M_{string}$.
But on the other hand there
is likewise no reason that the gauge couplings
will meet at some intermediate scale if they
are split by the thresholds at the string scale.
Below we shall give arguments that
there are stringy threshold corrections which
conserve gauge coupling unification for all group factors present
in a certain model.

There has been considerable progress in the computation of threshold
corrections in string theories \cite{kap88}. In particular explicit results
have been obtained in heterotic string models based on orbifolds
\cite{thr00,may93}
and fermionic \cite{ant91.2} constructions as well as in type II superstrings
\cite{dol92}. Surprisingly it was found that generically the threshold effects
cause only a minor change of the unification scale $M_X$ smaller than 5\%
\cite{kap88,dol92}.
This was unexpected because there are thousands of heavy states
which could a priori contribute huge corrections since they are not
arranged in representations of a unifying gauge group. Therefore
these string constructions involve
remarkably mild threshold effects. There are
two exceptions of this proper behavior. First the fermionic model
of ref. \cite{ant91.2}. Secondly the moduli
dependent threshold corrections of ref. \cite{thr00}, present in certain
orbifold models can grow even larger for sizeable VEV's of the
moduli fields. This is because states with non--trivial windings and
momenta in the internal dimensions have masses depending on the shape
of the internal manifold which in turn is parametrized by the
moduli. For large VEV's of some modulus complete string mass levels
are pushed below the string scale and give rise to considerable
threshold effects. On the other hand these kind of corrections are
only present in orbifold models with so--called $N=2$ sectors as
was explained in refs. \cite{thr00}.

In the following we specialize to orbifold constructions \cite{orb00}.
It is important to note that the up to now discussed
threshold corrections
are limited to the case of models with a (2,2) supersymmetry on the
worldsheet and a symmetry group $SL(2,Z)$ for the duality transformations
on the moduli manifold. However it is known that the discrete symmetry
group $\Gamma$ acting in the moduli space is generically different
from $SL(2,Z)$ as was shown in \cite{can91} for a certain Calabi--Yau
compactification and in \cite{erl92} for a special \zd orbifold. Recently
the computation of the moduli dependent threshold corrections was
extended to a large class of orbifold models with the result
that even in these rather simple toy models
the duality group $\Gamma$ can be at most a certain discrete
subgroup of $SL(2,Z)$ \cite{may93}.
These constructions are yet far from giving semi--realistic models
of our world since they have necessarily large gauge groups like
$E_6$ or $E_7$ and an unrealistic particle spectrum. To get smaller
gauge groups  one introduces background gauge fields corresponding
to Wilson lines with abelian \cite{iba87.1} or non--abelian \cite{iba87.2}
embedding
of the space group into the gauge degrees of freedom. In addition
one chooses a non--standard embedding of the orbifold twist
and as a consequence these models will possess only  (2,0) supersymmetry
on the worldsheet. New moduli fields, defined in general as marginal
deformations of the conformal field theory describing the two--dimensional
string sigma model, will give rise to additional threshold effects
parametrized by the VEV's of these fields.

In general it is quite complicated to calculate these corrections
because of technical difficulties. While it is relatively easy to
derive a formal expression for them there is no known general
way to perform the final $\tau$--integration over the modular parameter
of the torus. Still there remains the possibility to do a numerical
calculation. The disadvantage of this method is the difficulty to include
continuous deformations of the theory. Therefore, as a first step,
we focus on discrete changes in the background values of certain gauge
fields. Of course this choice will exclude the possibility of
getting large threshold corrections by the choice of appropriate large
VEV's for some Wilson line moduli as it was the case for the moduli dependent
threshold corrections. Nevertheless our model has at least a reasonable
gauge group structure and will provide some knowledge about the
situation in more realistic models.

The partition function for the \zd orbifold with discrete Wilson lines
is given by
\be
Z = \frac{1}{N} \left[ Z_{0,0} + \sum_{m,n} \eta(n,m)Z_{m,n}
\right] \ ,
\ee
where $Z_{0,0}$ is the partition function of the torus \cite{nar87}
and $Z_{m,n}$ is that of a sector twisted by $\Theta^m$ ($\Theta^n$)
in the $\sigma$ ($\tau$) direction of the torus:
\bea \label{twp}
Z_{m,n} &=& \int_{\tau \in \Gamma}\frac{d^2\tau}{\tau_2^2}
\frac{1}{2\pi\tau_2}\ Z_L(m,n) \ \bar{Z}_R(m,n)\ \bar{C}(m,n,\xi) \nn \\
&&\times \ds{\sum_{v^\prime \in \hat{v}}} \tilde{C}(m,n,\xi,\hat{v}) \
Z_{E_8}(m,n,v^\prime_1) \ Z_{E_8}(m,n,v^\prime_2) \ ,
\eea
where
\bea
Z_L(m,n) &=& \frac{\eta}{\ds{\prod_{\alpha = 2}^{4}}
\vartheta \left[
\ba{l}
1/2 + m\  \xi_\alpha \\
1/2 + n\  \xi_\alpha
\ea \right]} \nn \; , \\
\vspace*{1cm}
\bar{Z}_R(m,n) &=& \frac{\ds{\sum_{\alpha,\beta}}\eta_{\alpha\beta}\
\ds{\prod_{\alpha = 1}^{4}}
\bar{\vartheta} \left[
\ba{l}
\alpha + m\  \xi_\alpha \\
\beta + n\  \xi_\alpha
\ea \right]}
{\ds{\prod_{\alpha = 2}^{4}}
\bar{\vartheta} \left[
\ba{l}
1/2 + m\  \xi_\alpha \\
1/2 + n\  \xi_\alpha
\ea \right]\ \bar{\eta}^3} \; ,\\
Z_{E_8}(m,n,v) &=& \frac{
\ds{\sum_{\alpha,\beta}}\eta_{\alpha\beta}(m,n)\
\ds{\prod_{I = 1}^{8}}
\vartheta \left[
\ba{l}
\alpha + m\  v^I \\
\beta + n\  v^I
\ea \right]}{\eta^8} \ .\nn
\eea
Here $\tau = \tau_1 + i \tau_2$
is the single Teichm\"uller parameter of the torus,
$\bar{C}(m,n,\xi)$ and $\tilde{C}(m,n,\xi,\hat{v})$ are numerical factors
which take care of the degeneracy of the fixed points \cite{iba88,fon90},
$\eta(m,n)$, $\eta_{\alpha\beta}$ and $\eta_{\alpha\beta}(m,n)$ are
phase factors determined from modular invariance and $\vartheta$ and
$\eta$ are Jacobi and Dedekind functions.
$\xi_\alpha = 1/3(0,1,1,-2)$
 is the shift vector describing the boundary conditions
of the space--time degrees of freedom and the shift $v^{\prime I}$
in the gauge sector has been split into two parts, $v^\prime_1$,
$v_2^\prime$ acting in the two
$E_8$ factors. The sum in eq. (\ref{twp}) runs over
the following set of shift vectors:
\be
\hat{v} = \left\{ v^\prime:\ v^\prime =
 v + \ds{\sum_{k=1}^\omega} s^k a_k^I,\ s^k \in \{0,1,-1\} \right\} \ ,
\ee
where $\omega$ is the number of independent Wilson lines. Note that
in general the phases inside a $E_8$ factor depend on the sector
$(\Theta^m,\Theta^n)$.

The expressions given above apply immediately
also to other ${\bf Z}_N$ orbifolds if they are supplemented
by the contributions of the internal zero--modes in sectors with fixed tori
as was explained in ref. \cite{erl93}.

The threshold corrections are given by a slight deformation of the
above partition function \cite{kap88}:
\be \label{thre}
\btu_a = \ds{\int_{\tau \in \Gamma}} \frac{d^2\tau}{\tau_2}
\left( {\cal B}_a(\tau,\tb) - b_a \right) \ ,
\ee
where $ b_a = {\cal B}_a(\tau = i\infty) $. Apart from factors of $\tau_2$,
${\cal B}_a$ differs from the previously
given partition function only by insertions of two kinds of charge operators.
One of them acts in the fermionic space--time degrees of freedom
and is sensitive to the Lorentz representation of a certain state.
The other
one is inserted in the gauge degrees of freedom and counts the
charge of the state with respect to a specific
Cartan generator of the gauge group.
Both insertions can be taken into account by the use of the following
modified $\vartheta$ functions:
\bea
\vartheta^{\prime\prime}\left[
\ba{l}
\theta \\
\phi
\ea \right](\tau) &=& \ds{\sum_{n}}(n+\theta)^2
q^{(n+\theta)^2/2}e^{2\pi i(n+\theta)\phi} \nn\ , \\
\vartheta^\prime\left[
\ba{l}
\theta \\
\phi
\ea \right](\tau) &=& \ds{\sum_{n}}(n+\theta)
q^{(n+\theta)^2/2}e^{2\pi i(n+\theta)\phi}\ .
\eea
In general the insertion of the charge operator with respect
to a certain Cartan
generator requires the replacement of a factor $Z_{E_8}$ in the
partition function by a linear combination of deformed
$Z_{E_8}$ factors with the above modifications in the appropriate
$\vartheta$ functions. The concrete expression can be most easily
obtained in a basis of the gauge lattice where the  non--abelian
gauge groups are in one--to--one correspondence with definite
$\vartheta$ functions.

As a concrete model we consider the
\zd orbifold with gauge group $[SU(3)\times SU(2) \times U(1)^5]\times
[SO(10)\times U(1)^3]^\prime$
constructed in ref. \cite{iba87.3}
and further elaborated in ref. \cite{cas89}. The model is defined by the
shift vector $v$ and two Wilson lines $a_1^I$ and $a_3^I$:
\bea
v^I &=& \frac{1}{3}(1\ 1\ 1\ 1\ 2\ 0\ 0\ 0)(2\ 0\ 0\ 0\ 0\ 0\ 0\ 0) \nn \ ,\\
a_1^I &=& \frac{1}{3} (0\ 0\ 0\ 0\ 0\ 0\ 0\ 2)(0\ 1\ 1\ 0\ 0\ 0\ 0\ 0) \ ,\\
a_3^I &=& \frac{1}{3} (1\ 1\ 1\ 2\ 1\ 0\ 1\ 1)(1\ 1\ 0\ 0\ 0\ 0\ 0\ 0)\nn\ .
\eea
The massless spectrum of this model was given in detail in ref.
\cite{cas89}.

We computed numerically the threshold corrections with respect to
the gauge group factors $SU(3),\ SU(2),\ SO(10)$ and one $U(1)$. The $U(1)$
we have chosen is that one denoted by $U_5$ in \cite{cas89}.
A special property of eq. (\ref{thre})
is the fact
that it applies not to a single threshold correction to a gauge coupling
but only to the difference of such corrections w.r.t. two different
gauge couplings.
Therefore, to be precise,
we have computed in this way the relative threshold corrections
$\btu_{a-b}\equiv \btu_a-\btu_b$ .
The
coefficients $b_a$ of the $\beta$--function are
\be
b_{SU(3)} = 9, \hspace{1cm} b_{SU(2)} = 18,
\hspace{1cm}b_{SO(10)} = -18, \hspace{1cm}
b_{U(1)} = 32 \ ,
\ee
where we have normalized the quadratic Casimir of the gauge group as
tr($Q_a^2$)= 1/2 (1) for SU(N) (SO(N)).

To minimize the numerical errors which could possibly
arise because of cancellations of large terms,
we first calculated analytically the power series in
$e^{-\pi \tau_2}$  of the difference
of the modified partition functions.
The rapid convergence of this series can be easily checked with the standard
methods of analysis. Afterwards we integrated this series numerically
over the fundamental region $\Gamma$.

We found the surprising result that all relative threshold corrections
are identical after
division by the difference of the $\beta$-function coefficients:
\be \label{dda}
\btu \equiv \frac{\btu_a-\btu_b}{b_a-b_b} \approx 0.079 \ .
\ee
Remarkably this result implies that the running couplings
of all gauge groups meet in one point in spite of the fact that they are
no longer equal at the string scale $M_{string}$.
To gain insight into the structure of the threshold corrections we
make the following ansatz:
\be \label{watg}
\btu_a = b_a \cdot \btu + k_a Y \ , \; \; k_a = 1 \ .
\ee
This relation is motivated by the above mentioned fact that the formula
(\ref{thre}) determines a single threshold correction only up to
a group independent additive term $Y$. Checking the expressions for
the separate threshold corrections $\btu_a$ for $a \in \{SU(3),SU(2)
,U(1)_5,SO(10)\}$ we find that  ansatz (\ref{watg}) is justified.
We are then able to determine the value of the gauge group
independent term to be
\be
Y \approx 4.41 \ .
\ee
The outcome that the single threshold corrections $\btu_a$ can be written
as in eq. (\ref{watg}) is a highly non--trivial one,
the general form being $\btu_a = \tilde{\btu}_a + k_a Y$.
Where does this result come from ? The history of string
theory shows that modular invariance is quite often the root
of such surprising relations. Indeed also in computations
of moduli dependent threshold corrections \cite{thr00,may93} this
symmetry was responsible for the fact that the final expression
appeared to be a $\tau$--integral over the fundamental region
multiplied
by the difference of the $\beta$--function coefficients of the
$N=2$ supersymmetric sector of the theory. Therefore modular
invariance could be a plausible candidate to explain this
feature of the relative threshold corrections and might be a hint for
an analytic calculation of this quantity in $N=1$ supersymmetric sectors
as well as in models with Wilson lines.

Another peculiarity of the quantity $\btu$ in (\ref{dda}),
which is the relevant one for the discussion of  gauge coupling
unification, is its size.
Indeed it was found in ref. \cite{kap88}
 that for the standard \zd orbifold with gauge
group $E_8\times SU(3) \times E_6$ the relative threshold correction
w.r.t. the gauge groups $SU(3)$ and $E_6$ is zero while that w.r.t.
$E_8$ and $SU(3)$ results in a correction $\btu$ almost
equal to that found in our complicated model.
Moreover, for the \zd orbifold
with non--standard embedding and gauge group $[E_6 \times SU(3)]^2$ the
relative threshold corrections $\btu_{E_6-SU(3)}$ yield a $\btu$
of the same size as that of the standard ${\bf Z}_3$ orbifold.

To investigate further this issue we analyzed two additional
\zd orbifolds, the standard \zd orbifold with gauge group
$E_8\times SU(3) \times E_6$ mentioned above as well as a \zd orbifold
with one Wilson line and gauge group $[E_6 \times SU(3)]^2$ constructed
in ref. \cite{iba87.1}. Note that this second model has nothing to do
with the \zd orbifold of ref. \cite{kap88}
with the same gauge group but constructed without a
Wilson line. First it should be regarded as the
standard \zd orbifold with a non--vanishing vacuum expectation value
for certain background gauge fields. Secondly the particle spectrum is
different from that of the model without a Wilson line.

As a result we got that the power series which describes the relative
threshold corrections before integrating over the fundamental
region starts with the same term in all models under consideration
after division by the difference of the coefficients $b_a$.
This is not true for the terms of  higher order in $e^{-\pi \tau_2}$
which depend on the specific model. Since the series converges
very rapidly these differences cause only minor changes in the
total value of the threshold corrections. We have no explanation
for this agreement but it might well be that this common
first term is a reflection of the fact that all our models are
based on the \zd orbifold.

The threshold correction (\ref{dda}) amounts to a small increase in the
effective unification scale $M_X$ of about 4\%. Apart from the fact
that the shift of $M_X$ goes in the opposite direction, its size is
too small to explain the missing factor of
25 between $M_X$ and $M_{string}$.
On the other hand it is comparable
in size to the threshold corrections which arise from the shift $v$
associated to the twist $\Theta$ \cite{kap88}.
Since the Wilson line appears also in the internal momenta in sectors
with extended supersymmetry, quite similar to the moduli fields describing
the shape of the internal manifold, the natural size of the Wilson line
dependent threshold corrections from these sectors should be comparable to
that of the moduli dependent ones. Moreover we have seen from the
two standard \zd models -- with and without Wilson line -- that
the Wilson line dependent threshold corrections give {\em additional}
contributions: the relative thresholds $\btu_{E_6-SU(3)}$
of the standard \zd with vanishing background values for the gauge
fields are zero, while they are not for a non--trivial background
configuration.

Let us discuss now
the prospect of orbifold models to give proper string unification
compatible with the phenomenological observations.
To do this one has
of course to take into account all kinds of (threshold) effects
which influence the running of the gauge couplings. A classification
of orbifold models w.r.t. consistency with experiments was done in refs.
\cite{iba91}. The basic assumption was the existence of an orbifold model
with standard model gauge group and the massless particle content
of the MSSM. Subsequently the properties of such a hypothetical
model were discussed in the context ${\bf Z}_N$ and ${\bf Z}_N \times
{\bf Z}_M$ \cite{fon89} orbifolds. In this way a large class of orbifolds
have been ruled out to give phenomenological interesting models.

On the other hand the authors of refs. \cite{iba91} considered only threshold
corrections calculated for (2,2) models
and thus unbroken $E_6$ gauge symmetry. In contrast an orbifold model
with standard model gauge group is believed to be necessarily
a (2,0) model including an appropriate choice of Wilson lines.
Our above reasonings show that there will be additional threshold effects.
In particular these corrections should become sizeable in models with $N=2$
supersymmetric sectors. Since the expression for the Wilson line dependent
threshold corrections coming from these sectors is not known,
no statement can be made about their possible effects.
Moreover,
one has to take into account the mechanism which lowers the rank of the
gauge group. Continuous Wilson lines might be one possibility, but a
more general mechanism might also involve
the anomalous
$U(1)$ factor present in many (2,0) models \cite{fon89.2,cas89}. The anomalous
$U(1)$
will be broken \cite{din87} and during
this process otherwise massless particles
might acquire  a nontrivial mass that depends on the details
of the model. These states will heavily change the evolution of the
gauge couplings and indeed it was shown in refs. \cite{add00}
that the experimental requirements
can be met if certain conditions on their mass spectrum are fulfilled.
Note that the
effective massless particle content below
the mass scale of these states will
 be given by that of the MSSM.

Indeed, one might ask the question, whether a consistent and realistic
string model  exists in which the states lighter than the Planck mass
are exactly those of MSSM.
In fact the additional "massless" states usually present in orbifold
models are quite appropriate to play the role of particles at  an intermediate
mass scale as considered in refs. \cite{add00}. Therefore a reasonable
physics requirement for a model might be to ask for the massless particle
content of the MSSM at low energies and not at the string scale.

This has also consequences for the discussion of anomaly freedom of the
effective theory w.r.t. target space duality transformations on the
moduli space of the internal manifold. The anomaly coefficient
of the mixed gauge--duality anomalies depends on the properties of the
massless states under these transformations
which is described by the so--called modular weights.
In a completely twisted plane -- the so--called $N=1$ sector --
the anomaly coefficient has to be common to
all gauge groups to be cancelled
by a universal Green--Schwarz counterterm \cite{der91}. This gives a
condition on the modular weights of gauge group non--singlets.
These states, massive at an intermediate mass scale but massless
at the level of the string construction, contribute to the anomaly
coefficient with their
modular weights. Therefore one should not impose the condition
of anomaly freedom on the massless particle spectrum of the low energy
phase -- that of the MSSM -- but on the massless spectrum at the string scale
which includes the additional states.

There are two more general difficulties in the discussion of duality
anomalies. They both have to do with the definition of the modular
weights of the matter fields. The first is the exact form (and
knowledge) of the quantum symmetry group $\Gamma$ itself. As already
mentioned $\Gamma$ is given by $SL(2,Z)$
only in the simplest models\cite{erl92,may93}.
Therefore it is not known whether the above concept  can be applied
to more sophisticated models involving non--standard embedding
and a set of continuous and/or discrete Wilson lines,
where $\Gamma$ is not explicitly known
and might be inappropriate for a useful definition of the modular
weights. The second difficulty in the definition of the modular weights
is the explicit knowledge of the moduli dependent \kae matter metric
$K_M(T_i,\bar{T_i})$, where $T_i$ denotes a chiral superfield containing a
modulus of the internal manifold. All derivations of this function
are based on a (2,2) supersymmetry on the worldsheet \cite{kam00}.
Therefore the
application of these results to (2,0) models has to be noticed
as an important assumption. Nevertheless interesting results have
been obtained in
refs. \cite{iba91} using naively the matter metric derived for (2,2)
models.\\

We have thus seen that the discussion of the properties of realistic
string models is far from being conclusive. The same
can be said about the interpretation of the apparent unification
of gauge coupling constants. In its simplest form it would favor a
standard GUT with $M_X\geq 10^{16}$GeV. Realistic versions of such
grand unified theories, however, require many heavy particles and
threshold effects due to such particles could change the successful
predictions. In addition $M_X$ seems to be quite close to the
Planck scale and the influence of the gravitational
interactions could become important. One might therefore consider
gauge coupling unification in string theory as a possible alternative
scenario. Also there we have to deal with many heavy states and
threshold effects. One interesting aspect of string unification is the fact
that it does not require a unified gauge group. Unification of
gauge coupling constants at the tree level is a universal property of
any string model, with or without a grand unified group, due to the
universal coupling of the dilaton. This fact, however, could be changed
by quantum effects and one might worry whether realistic string models
do predict gauge coupling unification. The results of the
calculation presented in this paper indicate a more general validity
of the tree level result. Corrections are mild and seem to effect
different gauge couplings in a similar way
(see eq. (11)). Of course, there still remains
the question concerning the unification scale as compared to the
string unified scale. This question, however, can probably only be answered
 with the knowledge of the heavy particle spectrum. A distinction
between the two scenarios from a low energy point of view could come
from the investigation of Yukawa-coupling unification and its role
in MSSM. While in standard GUTS such a unification (e.g. bottom-tau
Yukawa coupling degeneracy at $M_X$) is in general expected, no such statement
can be made in string unified models, although the possibility cannot be
ruled out. Thus the situation remains inconclusive.\\ \\

{\bf Acknowledgements\hfill}\\
One of us (P.M.) would like to thank Albrecht Klemm for useful discussions.

\footnotesize
\renewcommand{\baselinestretch}{0.1}

\end{document}